# Gaussian-distribution analysis of transverse momentum spectra of $K_0^S$ and $K^{*0}$ in $\sqrt{s_{NN}} = 200$ GeV collisions

Zhi-hao Pan, Si-you Guo, and B. C. Li

*College of Physics and Electronic Engineering, Shanxi University, Taiyuan 030006, China*

**Abstract:** In this paper, we use the Gaussian distribution to analyze transverse momentum spectra of $K_0^S$ and $K^{*0}$ at midrapidity in $d$ +Au, Cu+Cu and $p + p$ Collisions at $\sqrt{s_{NN}} = 200$ GeV. The two-component Gaussian distribution can approximately agree with the experimental data and the three-component distribution can perfectly agree with the data in the given range. Moreover, there is a significantly positive correlation between the temperature and the collision centrality.

1. Introduction

Many final-state particles are produced in the high-energy nuclear collisions. Distribution study of particles produced in different kinds of collisions enables us to understand the collision process in more ways. In final state of the collision, particle properties are focused specifically on, such as transverse momentum, dihadron azimuthal correlations, pseudorapidity, and so on. Transverse momentum spectra of hadrons produced in proton and heavy-ion collisions at RHIC and LHC energies have been described successfully in non-extensive statistical mechanics.

In our previous work, we combined Tsallis statistics with a multisource thermal model to study pseudorapidity distributions of charged particles produced in high-energy $pp$ or $p\bar{p}$ collisions and high-energy nucleus-nucleus collisions. Recently, Tsallis statistics and Boltzmann statistics have been used to analyze the transverse momentum spectra in heavy-ion collisions at high energy. They can both extract the thermodynamics parameters of matter produced in the collision. In order to understand concretely the thermodynamics properties, this work compare the Tsallis distribution, Boltzmann distribution and Gaussian distribution adopted in the multisource thermal model by properties of transverse momentum spectra in the high-energy collisions.

**2. Tsallis distribution, Boltzmann distribution and Gaussian distribution in the multisource thermal model**

The collision is regarded to contain two processes, the soft process and the hard process. Generally, the two processes do not have enough coherence with each other so that a single distribution has difficulty in the reproduction of the experimental data over a lager range. We need the distribution composed of multiple components

$$\frac{dN}{dP_T} = \sum_i^n \omega_i l_i(C_i, T_i), \tag{1}$$

where $l_i(C_i, T_i)$ is the $i$th component of the distribution, $\omega_i$ indicates the rate percentage of $i$th components and $\sum_i^n \omega_i = 1$. In this paper, the two components and three components distribution will be adopted.

2.1 Tsallis distribution

In the Tsallis statistics, the momentum distribution is

$$\frac{d^3N}{d^3P} = \frac{gV}{(2\pi)^3}\left[1 + (q-1)\frac{E-\mu}{T}\right]^{-\frac{q}{q-1}}, \tag{2}$$

where $q$ is the non-equilibrium degree of the collision system, $T$ is the temperature and $g$ is the degeneracy degree. The parameters $p, V, E, \mu$ are the particle momentum, volume of the system, energy and the chemical potential, respectively. This distribution can also be written in terms of transvers mass $m_T$ and the rapidity $y$

$$\frac{d^2N}{dP_T dy} = gV\frac{P_T m_T \cosh y}{(2\pi)^2}\left[1 + (q-1)\frac{m_T \cosh y - \mu}{T}\right]^{-\frac{q}{q-1}}. \tag{3}$$

When $\mu = 0$ and $y = 0$, the distribution is

$$\left.\frac{d^2N}{dP_T dy}\right|_{y=0} = gV\frac{P_T m_T}{(2\pi)^2}\left[1 + (q-1)\frac{m_T}{T}\right]^{-\frac{q}{q-1}}. \tag{4}$$

2.2 Boltzmann distribution

For the Maxwell ideal gas, the momentum distribution is

$$\frac{dN}{Nd^3P} = \frac{P^2}{(m_0 kT)^{3/2}}\sqrt{\frac{2}{\pi}}\exp\left(-\frac{P^2}{2m_0 kT}\right). \tag{5}$$

The relativistic effects are considered in the high energy collision. The momentum distribution is

$$\frac{dN}{Nd^3P} = \frac{P^2}{m_0^2 kT K_2\left(\frac{m_0}{kT}\right)} \exp\left(-\frac{\sqrt{P^2+m_0^2}}{kT}\right), \tag{6}$$

where $K_2\left(\frac{m_0}{kT}\right)$ is the second-order modified Bessel function. Because the emission in the collision is isotropic, a two-component distribution is

$$\frac{dN}{NdP_T} = \omega B_1(C_1, T_1) + (1-\omega)B_1(C_2, T_2), \tag{7}$$

$$B_1(C_1, T_1) = C_1 P_T \exp\left(-\frac{\sqrt{P^2+m_0^2}}{kT_1}\right), \tag{8}$$

where $\omega$ is the rate percentage of the first component. T

2.3. Gaussian distribution

For $P_x$ and $P_y$, the form of Gaussian distribution is

$$f(P_{x,y}) = \frac{1}{\sqrt{2\pi T}} \exp\left(-\frac{P_{x,y}^2}{2T}\right). \tag{9}$$

The transverse momentum is

$$f(P_T) = \frac{1}{\sqrt{2\pi T}} \exp\left(-\frac{P_T^2}{2T}\right). \tag{10}$$

In the work, we use a two-component distribution and a three-component distribution. The $i$th component of the distribution $g_i(C_i, T_i)$ is

$$g_i(C_i, T_i) = \frac{C_i}{\sqrt{2\pi T_i}} \exp\left(-\frac{P_t^2}{2T_i}\right). \tag{11}$$

The two-component distribution is

$$\frac{dN}{dP_T} = C_0[\omega g_1(C_1, T_1) + (1-\omega)g_2(C_2, T_2)], \tag{12}$$

where $\omega$ and $1-\omega$ are the rate percentages of the two components. The three-component distribution is

$$\frac{dN}{dP_T} = C_0[\alpha g_1(C_1, T_1) + \beta g_1(C_1, T_1) + (1-\alpha-\beta)g_3(C_3, T_3)], \tag{13}$$

where $\alpha, \beta$ and $1-\alpha-\beta$ are the rate percentages of the three components.

## 3. Conclusion and discussion

Figure1 shows the results of the two-component Gaussian distribution corresponding to the transverse momentum spectra of $K_0^S$ mesons at mid rapidity in $d$+Au, Cu+Cu, and $p+p$ collisions at $\sqrt{s_{NN}} = 200$ GeV. The experimental data measured by STAR and PHENIX collaborations are shown with different symbols. For Cu+Cu and $d$+Au, different collision centralities are marked by the different shape symbols. At the bottom of the figure, we give the data of the $p+p$ collisions as a reference. It is clear that the two-component distribution agree mostly with the data, but it is slightly off the data when $P_T$ is less than about 3 GeV/c and exceeds about 11 GeV/c. The corresponding parameters used in the calculation are showed in Table 1. $T_1$ and $T_2$ indicate the two-component temperature. $\omega$ and $(1-\omega)$ are the rate percentage of the two components, respectively. $C_1$ and $C_2$ are coefficients of each component, $C_3$ is a total normalized coefficient. Then, we consider $C_1 = C_2 = 1$.

Figure3 shows the same data as Figure 1, but the model results are from a three component Gaussian distribution. It fits the data much better than the two-component distribution. It can agree with more experimental data. Parameters used in the calculation are given in the Table 3, where $T_0$, $T_1$ and $T_2$ indicate the three-component temperature. $\alpha$ and $\beta$ are rate percentages of the first two components. $C_1$, $C_2$ and $C_3$ are coefficients of each component. $C_0$ is the total normalized coefficient.

In Figure2 and Figure 4, we also give a comparison between the model results and the experimental points of $K^{*0}$ (or $\overline{K^{*0}}$). Similar to the case of $K_0^S$, Figure2 with the two-component Gaussian distribution does not agree well with the data at the low transverse momentum and high transverse momentum. The three-component Gaussian distribution behaves perfectly. Parameters used in the calculations are showed in the Table 2 and Table 4, where the parameter meaning is the same as that in Table 1 and Table 3.

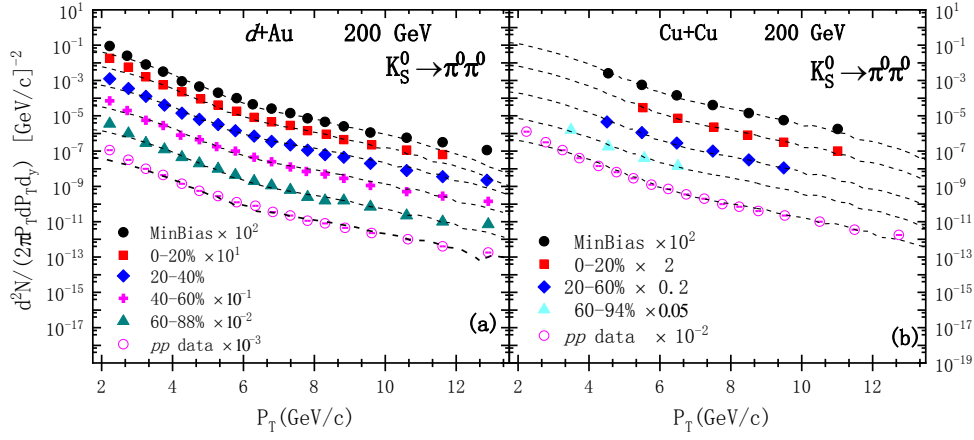

Figure 1 $K_0^S$ transverse momentum spectra in $d$+Au, Cu+Cu and $p+p$ collisions at $\sqrt{s_{NN}} = 200$ GeV. Experimental data [2-6] are shown with different symbols. The dashed lines are the results of Eq. (13).

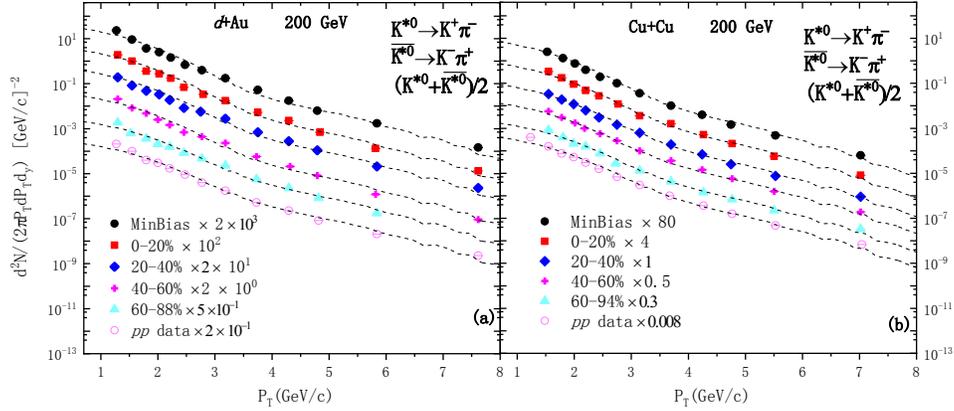

Figure 2 $K^{*0}$ and $\overline{K^{*0}}$ transverse momentum spectra in $d$+Au, Cu+Cu and $p+p$ collisions at $\sqrt{s_{NN}} = 200$ GeV. Experimental data [2-6] are shown with different symbols. The dashed lines are the results of Eq. (13).

.

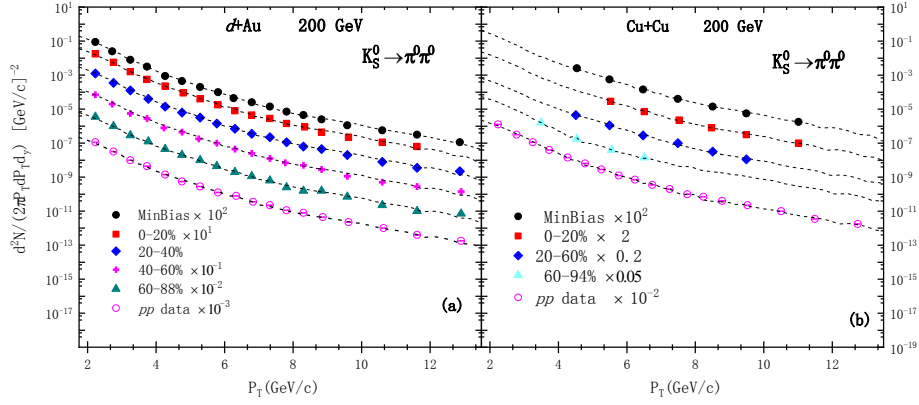

Figure 3 $K_0^S$ transverse momentum spectra in $d$+Au, Cu+Cu and $p+p$ collisions at $\sqrt{s_{NN}} = 200$ GeV. Experimental data [2-6] are shown with different symbols. The dashed lines are the results of Eq. (14).

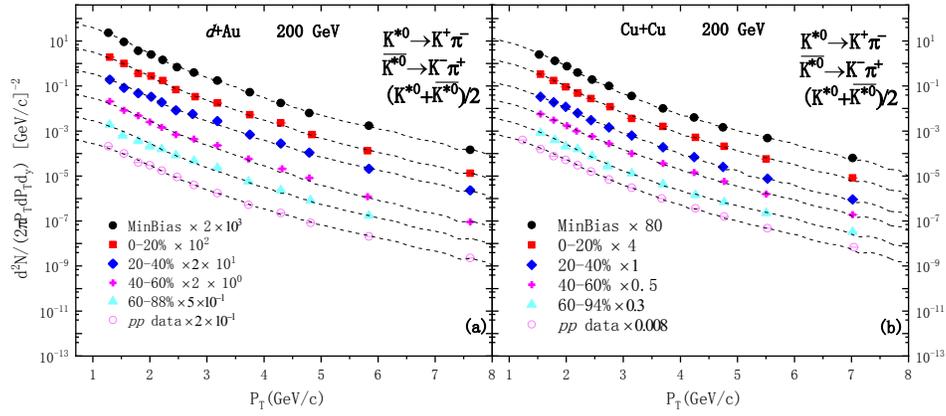

Figure 4 $K^{*0}$ and $\overline{K^{*0}}$ transverse momentum spectra in $d$+Au, Cu+Cu and $p+p$ collisions at $\sqrt{s_{NN}} = 200$ GeV. Experimental data [2-6] are shown with different symbols. The dashed lines are the results of Eq. (14).

Table 1 Parameters in figure 1.

| $d$+Au  Two component for $K_0^S$ | | | | |
|---|---|---|---|---|
| | $T_1$ | $T_2$ | $\omega$ | $C_0$ |
| MinBias | 2.04 | 9.40 | 0.977 | $6.302 \times 10^{-5}$ |
| $0-20\%$ | 2.08 | 9.50 | 0.977 | $1.087 \times 10^{-5}$ |
| $20-40\%$ | 2.10 | 9.60 | 0.971 | $8.572 \times 10^{-7}$ |
| $40-60\%$ | 2.13 | 9.95 | 0.972 | $5.065 \times 10^{-8}$ |
| $60-88\%$ | 2.16 | 10.20 | 0.9780 | $2.249 \times 10^{-9}$ |
| $pp$ | 2.19 | 10.50 | 0.969 | $6.076 \times 10^{-11}$ |
| Cu+Cu  Two component for $K_0^S$ | | | | |
| | $T_1$ | $T_2$ | $\omega$ | $C_0$ |
| MinBias | 2.22 | 9.11 | 0.976 | $1.980 \times 10^{-4}$ |
| $0-20\%$ | 2.30 | 9.23 | 0.973 | $1.075 \times 10^{-5}$ |
| $20-60\%$ | 2.32 | 9.40 | 0.973 | $3.279 \times 10^{-7}$ |
| $60-94\%$ | 2.38 | 9.26 | 0.954 | $1.243 \times 10^{-8}$ |
| $pp$ | 2.40 | 10.18 | 0.977 | $7.004 \times 10^{-10}$ |

Table 2 Parameters in figure 2.

| $d$+Au  Two component for $K^{*0}$ or $\overline{K^{*0}}$ | | | | |
|---|---|---|---|---|
| | $T_1$ | $T_2$ | $\omega$ | $C_0$ |
| MinBias | 0.85 | 3.94 | 0.964 | $5.89 \times 10^{-3}$ |
| $0-20\%$ | 0.86 | 3.96 | 0.962 | $6.35 \times 10^{-4}$ |
| $20-40\%$ | 0.85 | 3.965 | 0.959 | $8.00 \times 10^{-5}$ |
| $40-60\%$ | 0.86 | 3.968 | 0.959 | $6.00 \times 10^{-6}$ |
| $60-88\%$ | 0.91 | 3.99 | 0.936 | $4.90 \times 10^{-7}$ |
| $pp$ | 0.92 | 4.01 | 0.933 | $5.14 \times 10^{-8}$ |
| Cu+Cu  Two component for $K^{*0}$ or $\overline{K^{*0}}$ | | | | |
| | $T_1$ | $T_2$ | $\omega$ | $C_0$ |
| MinBias | 0.84 | 3.54 | 0.956 | $1.470 \times 10^{-3}$ |
| $0-20\%$ | 0.86 | 3.55 | 0.953 | $1.631 \times 10^{-4}$ |
| $20-40\%$ | 0.87 | 3.59 | 0.947 | $1.962 \times 10^{-5}$ |
| $40-60\%$ | 0.88 | 3.60 | 0.938 | $2.942 \times 10^{-6}$ |
| $60-94\%$ | 0.905 | 3.62 | 0.934 | $3.983 \times 10^{-7}$ |
| $pp$ | 0.91 | 3.63 | 0.943 | $9.310 \times 10^{-8}$ |

Table 3  Parameters in figure 3.

| d+Au  Three component for $K_0^S$ | | | | | | |
|---|---|---|---|---|---|---|
|  | $T_0$ | $T_1$ | $T_2$ | $a$ | $b$ | $C_0$ |
| MinBias | 1.06 | 3.18 | 12.18 | 0.902 | 0.095 | $1.904 \times 10^{-3}$ |
| $0-20\%$ | 1.06 | 3.19 | 12.30 | 0.889 | 0.107 | $3.559 \times 10^{-5}$ |
| $20-40\%$ | 1.07 | 3.22 | 12.38 | 0.901 | 0.095 | $2.940 \times 10^{-6}$ |
| $40-60\%$ | 1.08 | 3.25 | 12.52 | 0.884 | 0.112 | $1.528 \times 10^{-7}$ |
| $60-88\%$ | 1.08 | 3.30 | 12.60 | 0.883 | 0.112 | $6.278 \times 10^{-9}$ |
| $pp$ | 1.09 | 3.34 | 12.68 | 0.890 | 0.106 | $2.175 \times 10^{-10}$ |
| Cu+Cu  Three component for $K_0^S$ | | | | | | |
|  | $T_0$ | $T_1$ | $T_2$ | $a$ | $b$ | $C_0$ |
| MinBias | 1.05 | 3.53 | 12.40 | 0.874 | 0.121 | 0.000421 |
| $0-20\%$ | 1.11 | 3.60 | 12.70 | 0.881 | 0.114 | $2.31 \times 10^{-5}$ |
| $20-60\%$ | 1.15 | 3.64 | 12.95 | 0.850 | 0.144 | $6.745 \times 10^{-7}$ |
| $60-94\%$ | 1.20 | 3.70 | 13.25 | 0.630 | 0.368 | $8.061 \times 10^{-8}$ |
| $pp$ | 1.23 | 3.77 | 13.50 | 0.567 | 0.432 | $3.567 \times 10^{-9}$ |

Table 4   Parameters in figure 4.

| d+Au  Three component for $K^{*0}$ or $\overline{K^{*0}}$ | | | | | | |
|---|---|---|---|---|---|---|
|  | $T_0$ | $T_1$ | $T_2$ | $a$ | $b$ | $C_0$ |
| MinBias | 0.48 | 1.53 | 4.92 | 0.804 | 0.804 | $8.420 \times 10^{-3}$ |
| $0-20\%$ | 0.52 | 1.62 | 4.94 | 0.801 | 0.801 | $7.270 \times 10^{-4}$ |
| $20-40\%$ | 0.54 | 1.62 | 4.95 | 0.792 | 0.792 | $8.050 \times 10^{-3}$ |
| $40-60\%$ | 0.55 | 1.63 | 4.97 | 0.795 | 0.795 | $6.850 \times 10^{-6}$ |
| $60-88\%$ | 0.56 | 1.63 | 4.97 | 0.798 | 0.798 | $6.800 \times 10^{-7}$ |
| $pp$ | 0.58 | 1.64 | 4.99 | 0.794 | 0.794 | $6.800 \times 10^{-8}$ |
| Cu+Cu Three component for $K^{*0}$ or $\overline{K^{*0}}$ | | | | | | |
|  | $T_0$ | $T_1$ | $T_2$ | $a$ | $b$ | $C_0$ |
| MinBias | 0.56 | 1.45 | 4.40 | 0.855 | 0.130 | $1.940 \times 10^{-3}$ |
| $0-20\%$ | 0.57 | 1.48 | 4.45 | 0.826 | 0.159 | $2.372 \times 10^{-4}$ |
| $20-40\%$ | 0.58 | 1.48 | 4.46 | 0.769 | 0.215 | $2.417 \times 10^{-5}$ |
| $40-60\%$ | 0.57 | 1.50 | 4.50 | 0.761 | 0.216 | $4.065 \times 10^{-6}$ |
| $60-94\%$ | 0.58 | 1.53 | 4.55 | 0.840 | 0.139 | $6.150 \times 10^{-7}$ |
| $pp$ | 0.58 | 1.55 | 4.60 | 0.810 | 0.170 | $1.244 \times 10^{-7}$ |

In our previous work [1], we successfully reproduce the experimental results by the two-component and three-component Gaussian distribution. The two-component Gaussian distribution agrees mainly with the data, but the agreement is not better. The three-component Gaussian distribution can agree perfectly with the data in the given range. From the parameter tables, it is seen that the temperature and collision centrality have a positive relationship. Further work is required to find a function which can agree with experimental results in the larger range.